\newcommand{\cal}[1]{\mathcal {#1}}
\newcommand{\mib}[1]{\mathversion{bold}{#1}}
\newcommand{\wtilde}[1]{\widetilde{#1}} 

\def\bsub{\begin{subequations}}
\def\esub{\end{subequations}}
\def\beq{\begin{eqnarray}}
\def\eeq{\end{eqnarray}}
\def\bsub{\begin{subequations}}
\def\esub{\end{subequations}}
\def\b{\begin{equation}}
\def\bs{\begin{split}}
\def\es{\end{split}}
\def\e{\end{equation}}

\documentclass{PoS}

\usepackage{graphics} 

\title{Quark spin polarization and spontaneous magnetization in high density quark matter}

\ShortTitle{Quark spin polarization and spontaneous magnetization}

\author{\speaker{Yasuhiko Tsue}
\\
        Physics Division, Faculty of Science, Kochi University, Kochi 780-8520, Japan\\
        E-mail: \email{tsue@kochi-u.ac.jp}}

\author{Jo\~{a}o da Provid\^{e}ncia and Constan\c{c}a Provid\^{e}ncia\\
       CFisUC, Departamento de F\'{i}sica, Universidade de Coimbra, 3004-516 Coimbra, Portugal
}

\author{Hiroaki Matsuoka\\
        Graduate School of Integrated Arts and Science, Kochi University, Kochi 780-8520, Japan
}

\author{Masatoshi Yamamura\\
        Department of Pure and Applied Physics, Faculty of Engineering Science, Kansai University, Suita 564-8680, Japan
}

\author{Henrik Bohr\\
        Department of Physics, B.307, Danish Technical University, DK-2800 Lyngby, Denmark
}

\abstract{
By using the Nambu-Jona-Lasinio model with a tensor-type four-point interaction between quarks, 
it is shown that there exists a possibility of a spin polarized phase in quark matter at finite temperature and density.  
When there exists the spin polarization, the spontaneous magnetization may occur if the effect of the anomalous magnetic moment of quark is 
taken into account. 
An implication to the compact star objects with strong magnetic field is discussed when the spin polarization occurs.
}

\FullConference{The 26th International Nuclear Physics Conference\\
		11-16 September, 2016\\
		Adelaide, Australia}

\begin{document}

\section{Introduction}

In recent year, one of interests for physics governed by the quantum chromodynamics (QCD) may be to 
clarify the phase structure with respect to temperature, quark and/or isospin chemical potential, 
external magnetic field and so on \cite{Fukushima:2011}. 
In the region with low temperature and high baryon density, various phases are expected in quark matter, such as 
inhomogeneous chiral condensed phase \cite{Buballa:2015}, quarkyonic phase \cite{McLerran:2007}, 
color superconducting phase \cite{Alford:2008} and so forth. 
Recently, it has been shown that there is a possibility of the existence of spin polarized phase 
by using the Nambu-Jona-Lasinio (NJL) 
model \cite{Nambu:1961_1}, which is regarded as an effective model of QCD \cite{Klevansky:1992,Hatsuda:1994,Buballa:2005},  
with a tensor-type four-point interaction between quarks at zero temperature and finite baryon density \cite{Bohr:2012,Tsue:2012,Tsue:2013}.  
Also, under the external magnetic field, a possible QCD ground state 
has been investigated at zero and finite temperature in the case of one flavor \cite{Ferrer:2014}.

The investigation of a possible phase structure under the magnetic field is one of interesting subjects about QCD phase structure \cite{Anderson:2016}.   
The existence of a strong magnetic field is found in certain kinds of compact stars such as neutron star and magnetar \cite{Harding:2006}. 
Further, in the ultrarelativistic heavy-ion collision experiments, it is indicated that a strong magnetic field may be created in the early stage 
of nucleus-nucleus collisions \cite{Kharzeev:2008}. 

In this paper, we show that the spin polarized phase may be exist even at finite temperature in the region of high baryon density in quark matter 
due to the tensor-type four-point interaction between quarks \cite{Matsuoka:2016}. 
Further, it is shown that a spontaneous magnetization appears by introducing an external magnetic field, only if spin polarization occurs and the effect of 
an anomalous magnetic moment of quark is taken into account \cite{Tsue:2015_2}.

\section{Spin polarized phase originated from the tensor-type four-point interaction}

Let us start from the following two-flavor NJL model Lagrangian density with a tensor-type four-point interaction between quarks: 
\begin{eqnarray}\label{2-1}
{\cal L} &=&
\bar{\psi} i \gamma^{\rho} \partial_{\rho} \psi + 
G_S \left \{ (\bar{\psi} \psi)^2+(\bar{\psi} i \gamma^5 \vec{\tau} \psi)^2 \right \}\nonumber\\
& & 
-\frac{G_T}{4} \Big\{ (\bar{\psi} \gamma^\rho \gamma^\sigma \vec{\tau} \psi) \cdot 
(\bar{\psi} \gamma_\rho \gamma_\sigma \vec{\tau} \psi)  +(\bar{\psi} i \gamma^5 \gamma^\rho \gamma^\sigma \psi)
(\bar{\psi} i \gamma^5 \gamma_\rho \gamma_\sigma \psi) \Big \} \  , 
\end{eqnarray}
where $\psi$ represents the quark field with the two-flavor. 
Here, the first line represents the original NJL model Lagrangian density and the second line represents the tensor-type 
interaction introduced in this model. 
Under a mean field approximation, the above Lagrangian density is recast into  
\begin{eqnarray}\label{2-2}
{\cal L}_{\rm MFA}
={\bar \psi}\left(i\gamma^{\rho} \partial_{\rho} - M \right)\psi-F\left({\bar \psi}\Sigma_3 \psi\right)-\frac{M^2}{4G_S}-\frac{F^2}{2G_T} \ . 
\end{eqnarray}
Here, we define the dynamical quark mass $M$ with the chiral condensate $\langle {\bar \psi}\psi\rangle$ and the spin polarized condensate $F$, respectively, as  
\begin{eqnarray}\label{2-3}
& & M=-2G_S\langle{\bar \psi}\psi\rangle\ , \qquad
F=-G_T\langle{\bar \psi}\Sigma_3 \psi\rangle\ , \qquad
\Sigma_3=-i\gamma^1\gamma^2=\left(
\begin{array}{cc}
\sigma_3 & 0 \\
0 & \sigma_3
\end{array}
\right)\ ,
\eeq
where $\sigma_3$ is the third component of the Pauli spin matrix. 
We can easily derive the Hamiltonian density under the mean field approximation and by diagonalizing the Hamiltonian density, 
we can also derive the energy eigenvalues or single-particle energies of quark easily as  
\beq\label{2-5}
E_{{\mib p}}^{(\eta)}=\sqrt{p_3^2+\left(\sqrt{p_1^2+p_2^2+M^2}+\eta F\right)^2}\ , \qquad (\eta=\pm 1)
\eeq
where ${\mib p}=(p_1,p_2,p_3)$ is momentum.

\begin{table}[b]
\begin{center}
\begin{tabular}{ccc} \hline
$\Lambda$/GeV &  { $G_{{S}}$/{GeV}$^{-2}$} & { $G_{{T}}$/{GeV}$^{-2}$}\\ \hline\hline
0.631 & 5.5 & 11.0 \\
\hline
\end{tabular}
\caption{Parameters used here.}
\end{center}
\label{tab1}
\end{table}

The thermodynamic potential $\Omega$ at finite temperature $T$ and quark chemical potential $\mu$ can be expressed as 
\beq\label{2-6}
\Omega&=&
{\cal H}_{\rm MFA}-\mu ({\cal N}_P-{\cal N}_{AP})+{\cal H}_{\rm vacuum} -TS\ , 
\eeq
where
\beq\label{2-7}
& &{\cal H}_{\rm MFA}=\sum_{{\mib p},\eta,\tau,\alpha}E_{\mib p}^{(\eta)}\left(n_{\mib p}^{(\eta)}+{\bar n}_{\mib p}^{(\eta)}\right)
+\frac{M^2}{4G_S}+\frac{F^2}{2G_T}\ , \nonumber\\
& &{\cal H}_{\rm vacuum}=-\sum_{{\mib p},\eta,\tau,\alpha}E_{\mib p}^{(\eta)} \ , \qquad
{\cal N}_P=\sum_{{\mib p},\eta,\tau,\alpha}n_{\mib p}^{(\eta)}\ , \quad
{\cal N}_{AP}=\sum_{{\mib p},\eta,\tau,\alpha}{\bar n}_{\mib p}^{(\eta)}\ , 
\nonumber\\
& &
n_{\mib p}^{(\eta)}=\frac{1}{1+\exp\left(\left(E_{\mib p}^{(\eta)}-\mu\right)/T\right)}\ , \quad
{\bar n}_{\mib p}^{(\eta)}=\frac{1}{1+\exp\left(\left(E_{\mib p}^{(\eta)}+\mu\right)/T\right)}\ , \nonumber\\
& &S=-\sum_{{\mib p},\eta,\tau,\alpha}\Biggl[n_{\mib p}^{(\eta)}\ln n_{\mib p}^{(\eta)}+(1-n_{\mib p}^{(\eta)})\ln (1-n_{\mib p}^{(\eta)}) 
+{\bar n}_{\mib p}^{(\eta)}\ln {\bar n}_{\mib p}^{(\eta)}+(1-{\bar n}_{\mib p}^{(\eta)})\ln (1-{\bar n}_{\mib p}^{(\eta)})\Biggl]\ . \ \ \ \quad 
\eeq
Here, $\tau$ and $\alpha$ are indices for isospin and color of quark, respectively. 
Then, the thermodynamic potential can be reexpressed as 
\beq\label{2-8}
\Omega&=&-\sum_{{\mib p},\eta,\tau,\alpha}\left[
E_{\mib p}^{(\eta)}+T\ln\left(1+\exp\left(-\frac{E_{\mib p}^{(\eta)}-\mu}{T}\right)\right)
+T\ln\left(1+\exp\left(-\frac{E_{\mib p}^{(\eta)}+\mu}{T}\right)\right)\right] \nonumber\\
& &+\frac{M^2}{4G_S}+\frac{F^2}{2G_T}\ . 
\eeq
When the sum with respect to three-momentum ${\mib p}$ is replaced to the integration, we have to introduce the three-momentum cutoff $\Lambda$. 
Also, the isospin and color degrees of freedom give factor 2 and 3, respectively. 
As a result, the sum is replaced as following way: 
\beq
\sum_{{\mib p},\tau,\alpha}\longrightarrow \frac{3}{\pi^2}\int_0^{\Lambda}dp_T \int_0^{\sqrt{\Lambda^2-p_T^2}}dp_3\ p_T\ , 
\nonumber
\eeq
where $p_T=\sqrt{p_1^2+p_2^2}$.

\begin{figure}[t]
\centering
\includegraphics[height=13cm]{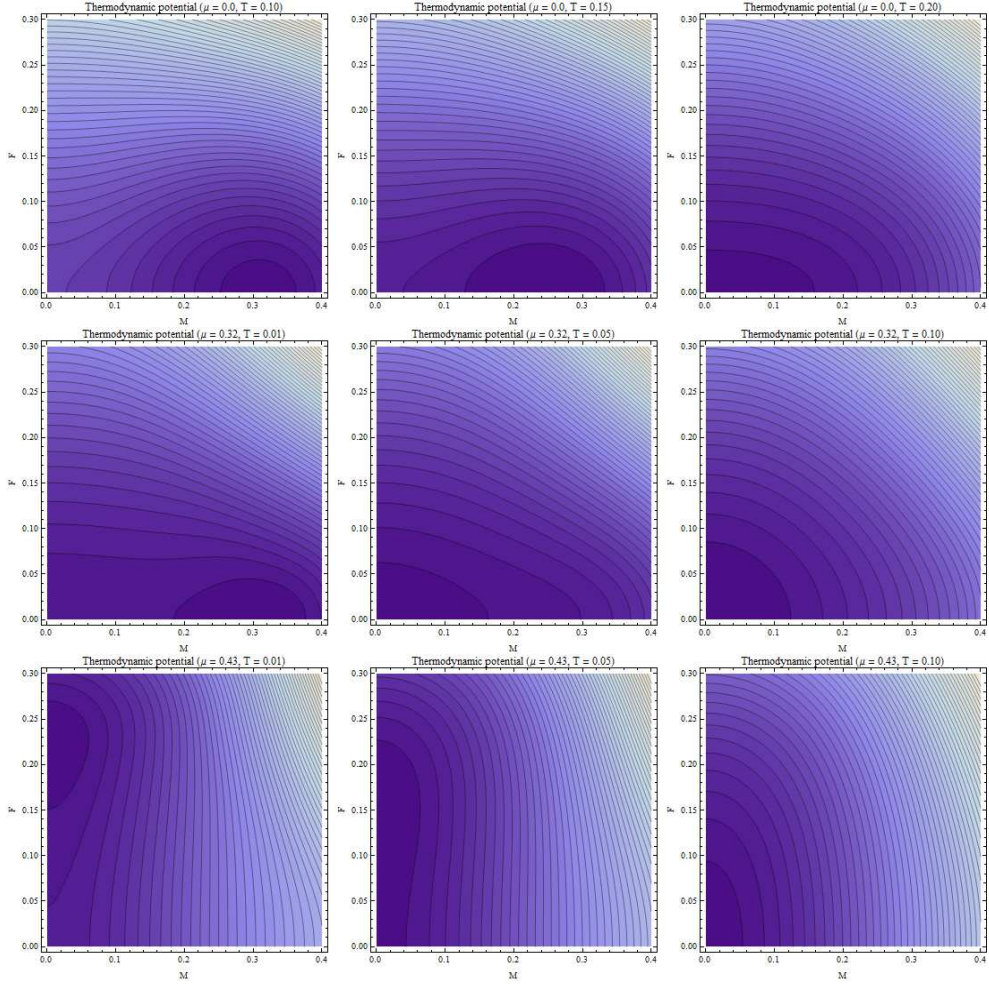}
\caption{The contour map of the thermodynamic potential is depicted. 
The horizontal and vertical axes represent the dynamical quark mass $M$ and the spin polarization $F$, respectively, 
with various quark chemical potential $\mu$ and temperature $T$.  
The darker color represents the lower value of the thermodynamic potential.}
\label{fig1}
\end{figure}

\begin{figure}[t]
\centering
\includegraphics[height=6cm]{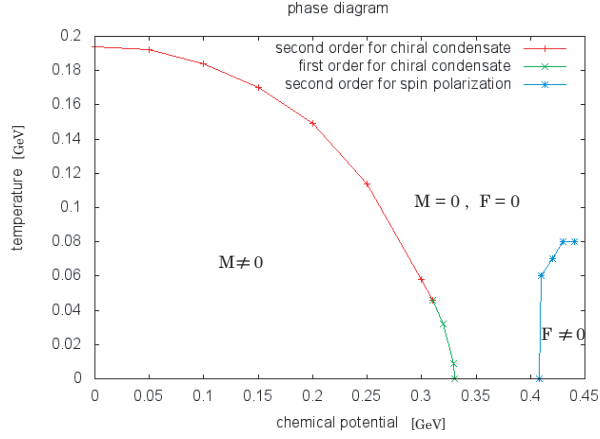}
\caption{The phase diagram is shown. The horizontal and vertical axes represent quark chemical potential and temperature, respectively. }
\label{fig2}
\end{figure}

We use the parameters in Table 1. 
The coupling constant $G_S$ and the three-momentum cutoff $\Lambda$ are determined by the quark condensate and the pion decay constant in 
vacuum. 
However, the coupling strength of the tensor-type interaction cannot be determined by the physical quantity in vacuum. 
In this paper, we regard $G_T$ as a free parameter.
It is shown in Figure 1 that chiral symmetry
is broken for small chemical potential and low temperature, namely the minimum point of the thermodynamic potential has 
$M\neq 0$ and $F=0$. 
However, if the chemical
potential or temperature becomes high, chiral symmetry is restored, namely, $M=0$ and $F=0$. 
In the large chemical
potential and low temperature region, the spin polarized condensate $F$ appears without $M$. 
However, for higher temperature, it disappears. 
In Figure 2,  the phase diagram in this model is presented on the $T$-$\mu$ plane, where 
the horizontal and vertical axes represent the quark chemical potential $\mu$ and temperature $T$, respectively. 
It is indicated that, for the boundary of the
chiral condensed and the normal phases, there is a critical endpoint for the phase transition
near $\mu = 0.31$ GeV and $T = 0.046$ GeV. 
On the other hand, the phase transition from the
normal quark phase to the spin polarized phase is always of second order.

\section{Spontaneous magnetization}

In this section, a possibility of spontaneous magnetization of quark matter is investigated by using the NJL model 
with the tensor-type four-point interaction between quarks. 
For this purpose, from beginning, the effect of the anomalous magnetic moment of quark, $\mu_A$, is introduced 
at the level of the mean field approximation. 
Here, we use a fact that the anomalous magnetic moment is expresses as \cite{AMM} 
\beq\label{6-1}
& &\mu_A{\bf 1}=
\left(
\begin{array}{cc}
\mu_u & 0 \\
0 & \mu_d
\end{array}
\right)\ , \qquad
\begin{array}{l}
\mu_u=1.85\ \mu_N\ , \\
\mu_d=-0.97\ \mu_N\ , 
\end{array}
\qquad
\mu_N=\frac{e}{2m_p}=3.15\times 10^{-17} \ {\rm GeV/T}\ . 
\eeq
We introduce the effects of the anomalous magnetic moment in the Lagrangian density within the mean field approximation.  
As a result, we obtain the following Lagrangian density :  
\beq\label{6-2}
{\cal L}_A={\cal L}_{\rm MFA}-\frac{i}{2}{\bar \psi}\mu_A\gamma^{\rho}\gamma^{\sigma}F_{\rho\sigma}\psi\ , 
\eeq
where $F_{\rho\sigma}=\partial_{\rho}A_{\sigma}-\partial_{\sigma}A_{\rho}$ and
$F_{12}\equiv -B_z=-B$. 
Here, ${\cal L}_{\rm MFA}$ is nothing but Eq.(\ref{2-2}) with $M=0$ because we consider the large chemical potential region. 
We only take $\rho$ and $\sigma$ as $\rho=1,\ \sigma=2$ and $\rho=2,\ \sigma=1$ because 
the magnetic field has only $z$-component.  
Then, in the mean field approximation, Eq.(\ref{6-2}) is recast into 
\beq\label{6-3}
{\cal L}_A=i{\bar \psi}\gamma^\rho D_{\rho}\psi 
-{\bar \psi}(F_3\tau_3+\mu_A B{\bf 1})\Sigma_3\psi-\frac{F^2}{2G_T}
=
i{\bar \psi}\gamma^{\rho}D_{\rho}\psi-{\bar \psi}{\wtilde F}\Sigma_3\psi
-\frac{F^2}{2G_T}\ .
\eeq
Here, since $F_3\tau_3=F{\bf 1}$ where ${\bf 1}$ is the $2 \times 2$ identity matrix for the isospin space, 
which are denoted in Eq.(\ref{2-3}), then 
we introduced the flavor-dependent variables ${\wtilde F}_f$ as 
\beq\label{6-4}
{\wtilde F}_f=F+\mu_f B\ , \quad{\rm namely}\quad 
{\wtilde F}_u=F+\mu_u B\ , \qquad
{\wtilde F}_d=F+\mu_d B\ .
\eeq
Thus, we can derive the thermodynamic potential in the same way developed 
in the section 2, except for the existence of the external magnetic field $B$. 
The energy eigenvalues are obtained as 
\beq\label{6-6}
E_{p_z,\nu,\eta}^{f}=\sqrt{p_3^2+\left({\wtilde F}_f+\eta\sqrt{2|Q_f|B\nu}\right)^2}\ , \qquad 
\left\{
\begin{array}{ll}
\nu=0,\ 1,\ 2, \cdots & {\rm for}\ \ \eta=1 \ , \\
\nu=1,\ 2,\ \cdots & {\rm for}\ \ \eta=-1\ , 
\end{array}
\right.
\eeq
with $Q_u=2e/3$ and $Q_d=-e/3$. 
Further, because the Landau quantization is carried out with respect to $p_T$, the integration with respect to $p_T$ is replaced to 
the sum with respect to integers $\nu$. 
As a result, the thermodynamic potential $\Phi$ at zero temperature with $M=0$ can be expressed as 
\beq\label{6-7}
\Phi=3\int_{-p_F}^{p_F}\frac{dp_3}{2\pi}\sum_{f=u,d; \eta=\pm}\frac{|Q_f|B}{2\pi}
\sum_{\nu=\nu_{\rm min}^{f\eta}}^{\nu_{\rm max}^{f\eta}}\left[E_{p_z,\nu,\eta}^f-\mu\right]+\frac{F^2}{2G_T}+({\rm vacuum\ term})\ , 
\eeq   
where $p_F$, $\nu_{\rm min}^{f\eta}$ and $\nu_{\rm max}^{f\eta}$ are determined by the condition $E_{p_z,\nu,\eta}^f \leq \mu$. 
Keeping in mind that we take $B\rightarrow 0$ finally in order to derive the spontaneous magnetization through the thermodynamic relation, 
the sum with respect to the Landau level $\nu$ can be replaced into the integration with respect to the continuum variable $a\nu$ with 
a small quantity $a$. 
A detailed derivation is found in Ref.\cite{Tsue:2015_2}. 
After the calculation, the results are summarized as follows: 
\beq\label{6-8}
& &\Phi=\Phi_>=-\frac{1}{2}\sum_{f=u,d}\frac{{\wtilde F}_f\mu^3}{2\pi}+\frac{F^2}{2G}+O(B^2)\ , \qquad\qquad\qquad
 {\rm for}\quad F>\mu\ , \nonumber\\
& &\Phi=\Phi_<=\frac{1}{2}\sum_{f=u,d}\frac{3}{\pi^2}\biggl[-\sqrt{\mu^2-{\wtilde F}_f^2}
\left(\frac{\mu^3}{6}+\frac{{\wtilde F}_f^2\mu}{4}\right)-
\frac{{\wtilde F}_f\mu^3}{3}\arctan\frac{{\wtilde F}_f}{\sqrt{\mu^2-{\wtilde F}_f^2}}
\nonumber\\
& &\qquad\qquad\qquad
+\frac{{\wtilde F}_f^4}{12}\ln \frac{\mu+\sqrt{\mu^2-{\wtilde F}_f^2}}{{\wtilde F}_f}\biggl]
+\frac{F^2}{2G} +O(B^2)\ , \qquad {\rm for}\quad 0 < F \leq \mu\ ,
\eeq
Thus, the spontaneous magnetization per unit volume, ${\cal M}$, can be 
derived thorough the thermodynamic relation, namely, 
${\cal M}=-\left.{\partial \Phi}/{\partial B}\right|_{B=0}
=-\sum_{f=u,d}\left.{\partial \Phi}/{\partial {\wtilde F}_f}\right|_{B=0}
\cdot{\partial {\wtilde F}_f}/{\partial B}$. 
From (\ref{6-8}), we can derive the spontaneous magnetization originated form the 
anomalous magnetic moment $\mu_u$, $\mu_d$ and the spin polarization $F$: 
\beq
{\cal M}&=&-\left.\frac{\partial \Phi_>}{\partial B}\right|_{B=0}
=\frac{\mu^3}{4\pi}(\mu_u+\mu_d)\ , 
\label{6-10}\\
{\cal M}&=&-\sum_{f=u,d}\left.\frac{\partial \Phi_<}{\partial {\wtilde F}_f}\right|_{B=0}
\cdot\frac{\partial {\wtilde F}_f}{\partial B}
=\frac{F}{2G}(\mu_u+\mu_d)\ , 
\label{6-11}
\eeq
where the gap equation $\partial \Phi/\partial F
=\sum_{f=u,d}{\partial \Phi}/\partial {\wtilde F}_f +\partial (F^2/(2G))/\partial F=0$ 
is used when Eq.(\ref{6-11}) is derived. 
It should be noted that if $F=0$, the spontaneous magnetization disappears because 
$F=0$ in Eq.(\ref{6-11}). 
\begin{figure}[t]
\centering
\includegraphics[height=4.5cm]{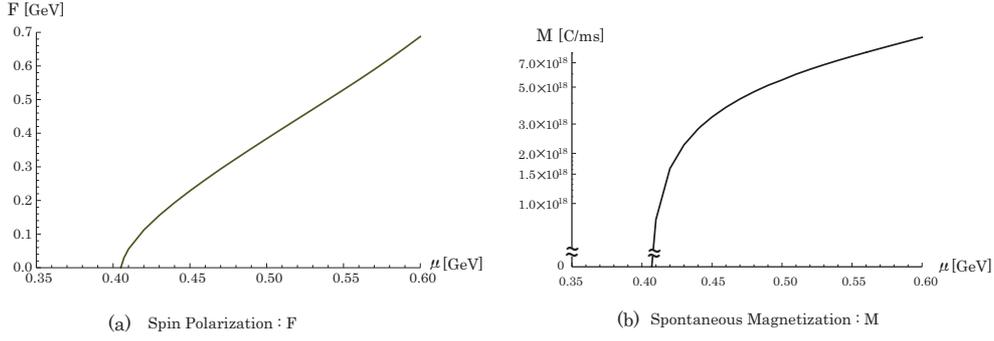}
\caption{(a) The spin polarization $F$ and (b) the spontaneous magnetization per unit volume ${\cal M}$ are depicted as functions with respect to 
the quark chemical potential $\mu$.}
\label{fig3}
\end{figure}
In Figure \ref{fig3}, the spin polarization $F$ and 
the spontaneous magnetization ${\cal M}$ are depicted as a function of the quark chemical potential $\mu$. 
For $\mu<\mu_c\approx 0.407$ GeV, the magnetization does not occur because the spin polarization does not appear, $F=0$. 
It is shown that, at $\mu=\mu_c$, the spontaneous magnetization appears connected with the spin polarization $F$.

\begin{figure}[b]
\centering
\includegraphics[height=4.3cm]{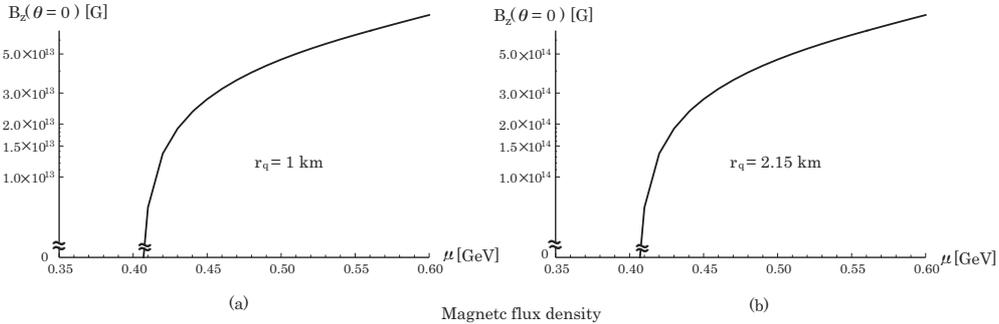}
\caption{The $z$-component of the magnetic flux density is shown in a logarithmic scale as a function with respect to the quark chemical potential $\mu$. 
(a) The case that the quark matter occupies 0.1 \% in the hybrid compact star and (b) the case that the quark matter occupies 1\%. .}
\label{fig4}
\end{figure}

Finally, let us roughly estimate the strength of the magnetic field due to the spontaneous magnetization.  
If the spontaneous magnetization per unit volume ${\cal M}$ can be regarded as the magnetic dipole moment, 
we can simply calculate the strength of the magnetic filed yielded by the spontaneous magnetization.
As is well known in the classical electromagnetism, 
the magnetic flux density ${\mib B}({\mib r})$ created by the magnetic dipole moment 
${\mib m}$ 
can be expressed as 
\beq\label{7-1}
{\mib B}=\frac{\mu_0}{4\pi}\left[-\frac{{\mib m}}{r^3}+\frac{3{\mib r}({\mib m}\cdot{\mib r})}{r^5}\right]\ , 
\eeq
where $\mu_0$ represents the vacuum permeability.
In our case, ${\mib m}=(0,0,{\cal M}\times V)$, where $V$ represents a volume. 
Here, let us consider the hybrid star with quark matter in the core of neutron star. 
Let us assume that the hybrid (neutron) star has radius $R=10$ km. 
If there exists quark matter in the inner core of star from the center to $r_q$ km, 
the strength of the magnetic flux density on the surface 
at the north or south pole of hybrid star is roughly estimated. 
Figure \ref{fig4} (a) and (b) 
shows the surface magnetic flux density as a function of the quark chemical potential $\mu$ 
in the case (a) $r_q=1$ km and (b) $r_q=2.15$ km where quark matter 
occupies (a) 0.1 \% and (b) 1 \% of the total volume of star. 
It is known that the strength of the magnetic field at the surface of magnetar is near $10^{15}$ G. 
Thus, in our calculation, if quark matter exists and the spin polarization occurs, 
a strong magnetic field about $10^{13}$ or $10^{14}$ Gauss may be created.

\section{Summary}

In this paper, it has been shown that a possibility of the spin polarized phase in quark matter at finite temperature and density 
is indicated by using the NJL model 
with the tensor-type four-point interaction between quarks as an effective model of QCD. 
When there exists the spin polarization, the spontaneous magnetization may appear if the effect of the anomalous magnetic moment of quark is 
taken into account. 
It is interesting to investigate what phase is realized under the external strong magnetic field created by the ultrarelativistic heavy-ion collision experiments and 
expected in the inner core of the compact stars such as neutron star and magnetar. 
The results of this subject will be reported elsewhere \cite{Tsue:2016}.

\section*{Acknowledgements}
One of the authors (Y.T.) 
is partially supported by the Grants-in-Aid of the Scientific Research 
(No.26400277) from the Ministry of Education, Culture, Sports, Science and 
Technology in Japan. 

\end{document}